# Van der Waals heterostructure mid-infrared emitters with electrically controllable polarization states and spectral characteristics

**Po-Liang Chen[1], Tian-Yun Chang[1], Pei-Sin Chen[1], Alvin Hsien-Yi Chan[2], Adzilah Shahna Rosyadi[2], Yen-Ju Lin[1, 3], Pei-Yu Huang[4], Jia-Xin Li[1], Wei-Qing Li[1], Chia-Jui Hsu[1], Neil Na[3], Yao-Chang Lee[4], Ching-Hwa Ho[2], Chang-Hua Liu[1, 5, 6]\***

1. Institute of Photonics Technologies, National Tsing Hua University, Hsinchu 30013, Taiwan

2. Graduate Institute of Applied Science and Technology, National Taiwan University of Science and Technology, Taipei 10607, Taiwan

3. Artilux Inc., Zhubei City, Hsinchu 30288, Taiwan

4. National Synchrotron Radiation Research Center, Hsinchu 30076, Taiwan

5. Department of Electrical Engineering, National Tsing Hua University, Hsinchu 30013, Taiwan

6. College of Semiconductor Research, National Tsing Hua University, Hsinchu 30013, Taiwan

**Abstract**

Modern infrared (IR) microscopy, communication, and sensing systems demand control of the spectral characteristics and polarization states of light. Typically, these systems require the cascading of multiple filters, polarization optics and rotating components to manipulate light, inevitably increasing their sizes and complexities. Here, we report two-terminal mid-infrared (mid-IR) emitters with electrically controllable spectral and polarization properties. Our devices are composed of two back-to-back p-n junctions formed by stacking anisotropic light-emitting materials, black phosphorus and black arsenic-phosphorus with $MoS_2$. By controlling the crystallographic orientations and engineering the band profile of heterostructures, the emissions of two junctions exhibit distinct spectral ranges and polarization directions; more importantly, these two electroluminescence (EL) units can be independently activated, depending on the polarity of the applied bias. Furthermore, we show that when operating our emitter under the polarity-switched pulse mode, its EL exhibits the characteristics of broad spectral coverage, encompassing the entire first mid-IR atmospheric window ($\lambda$: 3–5 µm), and electrically tuneable spectral shapes. Our results provide the basis for developing groundbreaking technology in the field of light emitters.

**Main text:**

The mid-infrared (mid-IR) spectral region is of great scientific and technical interest, primarily because this range of wavelengths falls within the molecular fingerprint region and covers two atmospheric windows ($\lambda$: 3–5 µm and 8–14 µm)[1,2]. To date, mid-IR light emitters have been widely used in diverse mid-IR imaging and spectroscopic systems, opening enormous opportunities for industrial, environmental, medical, defense and security sensing applications[3-5]. Continued efforts further demonstrate that these mid-IR systems can provide an additional dimension of contrast, such as revealing camouflaged surfaces or the compositional and functional properties of chemical species, if the polarization states and spectral characteristics of exploited mid-IR light can be actively controlled.[6] In addition to sensing applications, it is notable that mid-IR light sources are at the centre of modern free-space optical communication systems, and their information capacities can be further improved via the polarization- and spectral-encoding of optical signals[7,8]. However, despite promising advances, manipulating the spectral and polarization properties of mid-IR light generally requires the cascade of filters, dispersive optics and polarization optics together with mechanical moving parts. These mid-IR optical components are relatively more expensive and less developed than their visible or near-infrared counterparts. More critically, the requirements of multiple optical and mechanical elements pose fundamental limitations for realizing miniaturized systems with robust integration and high-speed operation.

To circumvent these technological obstacles, one possible approach is to develop mid-IR emitters with electrically controllable polarization and spectral properties. However, conventional mid-IR emitters generally rely on III–V or II–VI semiconductors[1,5]. Emissions from these three-dimensional materials are typically nonpolarized. Wavelength-tuneable emitters with electric control can in principle be achieved by vertically stacking two or more electroluminescence (EL) units (i.e., creating tandem structures). These emitters, operated within the visible region, have been successfully demonstrated using solution-processed organic polymers[9,10]. Unfortunately, these wavelength-tuneable devices require multilayered architectures to engineer the band profiles of heterostructures and multiple electrodes to independently control each EL unit. Exploiting epitaxial semiconductors, which generally suffer from lattice and thermal mismatches at heterointerfaces[1,5], to form tandem LEDs is a formidable challenge.

In these respects, exploiting a family of layered van der Waals (vdW) materials might be a promising alternative route for developing electrically controllable mid-IR sources. One prominent virtue offered by this material family is that different vdW materials can be stacked vertically with arbitrary chosen sequences and crystal orientations due to their weak vdW interactions[11-13]. Moreover, vdW materials provide a large variety of optical bandgaps, spanning across the electromagnetic spectrum from ultraviolet to terahertz[14,15]. For mid-IR emitter applications, black phosphorus (BP) has gained enormous attention, because it has a direct and narrow gap (~0.33 eV), low Auger recombination characteristics and an in-plane anisotropic structure[16-22]. By leveraging these exotic features, pioneering works have not only successfully demonstrated BP-based mid-IR LEDs[23-25], exhibiting characteristics of linear polarized emission and high emission efficiency, but have also shown their great promise for integrated mid-IR silicon photonics and gas sensing applications[23,26]. In addition, doping BP with arsenic to form black arsenic phosphorus (b-AsP) results in shrinkage in the band gap[16,20,27]. Applications of b-AsP to mid-IR photodetections with operational wavelengths greater than 8 µm have been demonstrated very recently[28,29]. However, the potential applications of using b-AsP on mid-IR emissions and on electrically tuneable sources remain experimentally unexplored.

In this article, we experimentally demonstrate mid-IR light emitters in the two-terminal configuration, in which their emission wavelengths and polarization states can vary with the polarity of the applied bias. The performed bias-switchable responses are made possible by assembling BP, b-AsP and $MoS_2$ together to create p-n-p junctions, which establish two distinct EL units, and by arranging their crystal orientations. In addition, our optoelectronic characterization indicates that the emission intensity of each EL unit can be electrically modulated with a radio frequency (RF) signal reaching a frequency of 3 MHz. By leveraging these features, we further demonstrate that applying two polarity-switched voltage pulse trains to the heterostructure emitter can lead to the generation of mid-IR light, showing the characteristics of a broad spectral range and tuneable spectral shape.

**Materials characterizations**

We start by characterizing the basic optoelectronic properties of the mid-IR building blocks used in our emitters. Figure 1a shows a schematic illustration of layered BP and b-AsP, which have orthorhombic lattices with puckered honeycomb structures. For b-AsP, we investigate two crystals with different As/P ratios: b-$As_{0.25}P_{0.75}$ and b-$As_{0.46}P_{0.54}$. The stoichiometries of these alloys are identified by Raman and energy-dispersive X-ray spectroscopy (EDX) measurements (see Supplementary Section 1). To examine their luminescence properties, we illuminated λ= 2.4 µm light onto the BP, b-$As_{0.25}P_{0.75}$ and b-$As_{0.46}P_{0.54}$ flakes, obtained by the mechanical exfoliation of synthetic crystals onto 285 nm $SiO_2$/Si substrates (Fig. 1b-d, inset), and then exploit the home-built mid-IR spectrometer to analyze their emission spectra (see Methods and Supplementary Section 2). As shown in Fig. 1b, the resolved photoluminescence (PL) spectrum of BP displays a fingerprinting peak at λ~3.7 µm; this value is close to the band gap energy of BP and consistent with previous reports[19,23]. However, the emission band maximum shifts to longer wavelengths as the As/P ratio in b-$As_xP_{1-x}$ increases (Fig. 1c-d). The observed PL shifts combined with our synchrotron-radiation-based Fourier-transform infrared (FTIR) spectroscopy measurements (Fig. 1e, see Methods) clearly indicate that alloying BP with As causes a reduction in the band gap. In addition to characterizing PL spectra, we experimentally verify that the emission of b-AsP shows a linear dichroism feature. Figure 1f displays the results of the polarization-resolved PL experiment performed on a b-$As_{0.46}P_{0.54}$ flake, which shows strong anisotropy with preferred linear polarization along the armchair crystal orientation of b-$As_{0.46}P_{0.54}$; this phenomenon is associated with the anisotropic selection rules near the band edge[16-22]. The degree of PL polarization of b-$As_{0.46}P_{0.54}$, defined as $\rho_{PL} = (I_{PL, max} - I_{PL, min})/ (I_{PL, max} + I_{PL, min})$, is determined to be 84%. Similar anisotropic optical transitions are found from BP and b-$As_{0.25}P_{0.75}$ flakes (see Supplementary Section 3), because they all share the same puckered structure. These results exemplify the possibilities of applying BP and b-AsP to linearly polarized mid-IR light-emitting diodes (LEDs).

## Mid-IR emitter with electrically controllable polarization states

Next, we describe the working principle of an electrically driven mid-IR vdW heterostructure emitter, in which its linear polarization emission is bias-switchable. Figure 2a,b presents a schematic and optical image of our proposed emitter, composed of MoS$_2$ (7.2-nm-thick) between two BP light-emitting layers. The top BP (BP$_t$) and bottom BP (BP$_b$) have thicknesses of 38 nm and 47 nm, respectively, and their armchair crystal orientations are orthogonal with each other (see Methods and Supplementary Section 4). Because BP has a smaller work function than MoS$_2$, the band alignment of the BP$_t$/MoS$_2$/BP$_b$ (top to bottom) heterostructures results in two rectifying junctions connected back-to-back; this finding is corroborated by our scanning photocurrent measurements (see Supplementary Section 5). In addition, it is notable that MoS$_2$ exhibits a large valence band discontinuity (ΔEv) but a nearly zero conduction band offset (ΔEc) with neighbouring BP[30,31]. Our finite element simulations, which are shown in Fig. 2c-d (also see Methods), indicate that such band alignment can cause electrons to flow unimpeded throughout the emitter, but the movement of holes is blocked by the barrier at the BP$_t$/MoS$_2$ (BP$_b$/MoS$_2$) interface when applying the positive (negative) bias voltage across two BP layers with BP$_b$ connected to the ground. Therefore, whether the recombination of electron-hole pairs occurs at BP$_t$ or BP$_b$ is associated with the polarity of the applied bias, offering the possibility for the electric control of the polarization direction of the emitted photons.

To verify our proposed device operation principle, we apply the bias voltage, $V_b$, onto the BP$_t$/MoS$_2$/BP$_b$ heterostructure emitter to examine its optoelectronic properties. Fig. 2e shows the measured $I$-$V_b$ characteristics. The rectification behaviour is observed at both positive and negative bias regions, which is correlated with the p-n-p junctions of the heterostructures. Interestingly, we find that such electrical excitation leads to mid-IR EL generation at both DC bias polarities (Fig. 2f-g). The peak of the measured EL spectrum is at λ= 3.67 μm, which is close to the PL result shown in Fig. 1b, and the peak wavelength remains unchanged with the variation of $V_b$ (see Supplementary Section 6), confirming that the emission originates from the electron-hole recombination in BP and not from black-body radiation. Moreover, our linear polarization-resolved measurement on EL reveals that the emission is strongly anisotropic (Fig. 2h). Under a positive (negative) bias, the EL intensity is maximized at $\theta = 90°$ ($\theta = 0°$), which is aligned with the armchair direction of BP$_t$ (BP$_b$). The calculated degree of EL polarization factors $\rho_{EL} = (I_{EL, max} - I_{EL, min})/(I_{EL, max} + I_{EL, min})$ are 88% for the positive bias and 87% for the negative bias. These features evidently confirm that a mid-IR emitter with electrically switchable and linearly polarized emission can be realized by leveraging the potentials of vdW heterostructures and the anisotropic properties of BP.

To gain further insight, we calibrate the bias-dependent output powers of the BP$_t$/MoS$_2$/BP$_b$ heterostructure emitter (see Supplementary Section 7 for the calibration). The results shown in Fig. 2i (black dots) indicates that its emission is initiated at low bias voltage ($|V_b| \sim 0.2$ V). With the increase in $V_b$, the output power approaches 0.5 μW based on the

~300 µm² light emitting area. From this result, we quantify the extrinsic quantum efficiency ($\eta$) by the following equation:

$$\eta = \frac{eN}{I}$$

where e is the electron charge; $N$ represents the number of photons emitted into the free space per second; and $I$ is the current injected into the emitter (see Supplementary Section 7 for more details). Figure 2i (brown dots) presents the bias-dependent extrinsic quantum efficiency (QE), which indicates that its peak efficiency can reach 0.3%. This performance is comparable to those of conventional mid-IR LEDs based on narrow gap compound semiconductors [1,32,33].

**Mid-IR emitter with electrically controllable polarization states and emission wavelengths**

In addition to the electric control of linear polarization, we elaborate that bias polarity can manifest the spectral response of heterostructure emitters. Figure 3a illustrates the schematic of our vdW-engineered emitter along with the optical image (Fig. 3b). The emitter consists of heterojunctions formed by vertical stacking of $BP/MoS_2/b-As_{0.46}P_{0.54}$ (top to bottom), in which 92-nm-thick $b-As_{0.46}P_{0.54}$ as well as 39-nm-thick BP serve as mid-IR light emitting layers, and the intermediate 7.9-nm-thick $MoS_2$ acts as a hole blocking layer. The armchair directions of BP and $b-As_{0.46}P_{0.54}$ are controlled to be orthogonal with each other during the stacking process (see Supplementary Section 8). Such a vdW stacking order leads to the formation of a p-n-p junction (see Supplementary Section 9). Thus, the emitter subjected to a positive (negative) bias leads to ambipolar injection with EL dominated by electron-hole recombination at the $BP/MoS_2$ ($b-As_{0.46}P_{0.54}/MoS_2$) interface, as illustrated in Fig. 3c-d (also see Methods), and its electrical transport shows the expected rectification behaviour at both bias polarities (Fig. 3e).

To confirm that our $BP/MoS_2/b-As_{0.46}P_{0.54}$ has bias-switchable features, we feed the generated EL into the mid-IR spectrometer. Figure 3f-g shows the representative EL spectra when the emitter is biased at $V_b$= 7 V and $V_b$= -10 V, respectively (see Supplementary Section 10 for EL spectra obtained at different $V_b$). The resolved EL characteristics, in terms of spectral shape and peak wavelength, clearly indicate that varying the bias polarity results in a change in the light emission from BP to $b-As_{0.46}P_{0.54}$. We note that such bias polarity-induced switching behaviour can be further corroborated by polarization-resolved measurements. As revealed in Fig. 3h, the emission of a positively (negatively) biased emitter is highly anisotropic with the linear polarization oriented along 90° (0°), corresponding to the armchair crystal orientation of BP ($b-As_{0.46}P_{0.54}$). These results indicate that $BP/MoS_2$ and $b-As_{0.46}P_{0.54}/MoS_2$ can be independently activated by tuning the bias polarity and that our proposed two-terminal heterostructures can be applied as a mid-IR emitter exhibiting bias-switchable polarization and spectral responses.

Next, we quantify the bias dependence of the emission power and extrinsic QE of the $BP/MoS_2/b-As_{0.46}P_{0.54}$ emitter by following the calibration procedure described in

Supplementary Section 7. Our analysis shows that the emission of the b-As$_{0.46}$P$_{0.54}$/MoS$_2$ EL unit is not as efficient as that of the BP/MoS$_2$ unit (Fig. 3i). This phenomenon might reflect the fact that b-As$_{0.46}$P$_{0.54}$ has a higher nonradiative Auger recombination rate ($\gamma_A$) than BP because $\gamma_A$ increases with decreasing band gap ($E_g$), as given by the equation[34]:

$$\gamma_A^{-1} \propto \exp\left[\left(\frac{m^*}{1+m^*}\right)\left(\frac{E_g}{K_B T}\right)\right]$$

where $m^*$ is the ratio of the electron and hole effective mass; $K_B$ is the Boltzmann's constant; and T is the temperature. Nevertheless, we find that the b-As$_{0.46}$P$_{0.54}$/MoS$_2$ EL unit can still emit up to 300 nW mid-IR light into free space. Additionally, the performance of the emitter, in terms of its emission power and EL spectral shape, remains unchanged after weeks of periodic measurements, suggesting the robustness of our developed technology and exploited vdW materials.

**Frequency responses and mid-IR broadband emission with electrically tuneable spectral shapes**

With sufficient power output, the frequency response of our BP/MoS$_2$/b-As$_{0.46}$P$_{0.54}$ emitter can be unambiguously resolved using a commercially available high-speed mid-IR InAsSb detector with a bandwidth of ~9 MHz and a limited responsivity of ~4 mA/W (see Methods). Figure 4a-b presents the variation in EL power when the applied bias voltage is positively (negatively) modulated to turn the BP/MoS$_2$ (b-As$_{0.46}$P$_{0.54}$/MoS$_2$) EL unit on and off at different frequencies. Notably, the modulation characteristic of the BP/MoS$_2$ (b-As$_{0.46}$P$_{0.54}$/MoS$_2$) unit exhibits a high 3-dB bandwidth of 3.8 MHz (5 MHz). These results imply that direct signal encoding without optical modulators and rapid switching between two distinct EL units can be performed.

Finally, we demonstrate that our BP/MoS$_2$/b-As$_{0.46}$P$_{0.54}$ heterostructure emitter can provide additional functionality when applying voltage pulse trains, possessing two pulses of opposite polarity in a cycle (Fig. 4c). Figure 4d presents the time-averaged EL spectrum when operating the emitter in pulse mode. The repetition rates of the pulse trains are all set at 1 kHz. Within each cycle, the positive and negative voltage pulses are set to $V^+$= 4 V and $V^-$= -6.5 V, respectively, but they both have the same duty cycle ($a/\tau = b/\tau = 15\%$). From this measurement, we find that the EL spectrum includes the luminescence features of BP and b-As$_{0.46}$P$_{0.54}$ (Fig. 4d), indicative of the excitation of both EL units under the polarity-switched pulse operation. Notably, the emission wavelength ranges from ~2.5 to 6 μm, covering the entire first mid-IR atmospheric window, whereas the spectral shape is dominated by BP due to the higher radiative decay rate of BP than b-As$_{0.46}$P$_{0.54}$. However, the more pronounced b-As$_{0.46}$P$_{0.54}$ emission can be observed by further increasing the amplitude of the negative voltage pulses (Fig. 4e-f), showing that the spectral shape is electrically tuneable.

**Conclusions**

In summary, this work provides the first demonstration of mid-IR emitters that can perform bias-switchable polarization and spectral properties. Our detailed characterizations reveal that they can exhibit high extrinsic QE and fast modulation speed. Central to these achievements is the combination of controlling the crystal orientations and band profiles of vdW heterostructures. In addition, we demonstrate that our developed heterostructures can be configured as mid-IR emitters, showing the characteristics of broad spectral coverage and electrically tuneable spectral shapes when operating under the polarity-switched pulse mode. The unique features (that is, electrically controllable polarization, emission wavelength and spectral shape) demonstrated here suggest that our emitters may have important implications in several advanced mid-IR sensing, data processing and imaging technologies[6-8,35,36]. Moreover, we envision that our developed technology can be extended to create other bias-switchable emitters operated within the near-IR or visible regions, considering that a variety of wide bandgap and anisotropic vdW light-emitting materials, such as $ReS_2$ and $TiS_3$, have been discovered in recent years[37,38].

**Methods:**

PL and EL measurements

The EL signals were generated by applying the bias voltage onto heterostructure emitters, and PL signals were obtained by illuminating BP or b-AsP with a $\lambda = 2.5$ μm laser beam. To analyze their spectra, the luminescence signals generated from the samples were collected by the objective (NA = 0.67) and sent into a self-manufactured mid-IR spectrometer equipped with a grating mounted on a rotation stage and a liquid nitrogen cooled InSb detector (see Supplementary Section 2 for the schematic diagram of the setup). All PL and EL experiments were conducted at room temperature; all light paths and the spectrometer were purged by passing high-purity nitrogen gas to minimize the spectral contribution of water vapour and $CO_2$.

Synchrotron-based FTIR spectroscopy

The absorption measurements of BP and b-AsP shown in Fig. 1e were conducted at the infrared microspectroscopy endstation TLS14A1 of the National Synchrotron Radiation Research Center (NSRRC). Specifically, the highly collimated synchrotron infrared beam was directed into a FTIR spectrometer (Thermo Scientific Nicolet 6700) and a confocal infrared microscope (Nicolet Continuum; ThermoFisher Scientific). The microscope was equipped with a 32x Cassegrain objective, which could focus the modulated infrared beam onto the BP or b-AsP samples with spot sizes of $10 \times 10$ μm$^2$. The reflected light from the sample was collected by the same objective and directed to a liquid nitrogen-cooled mercury-cadmium-telluride (MCT) detector. A computer then performed a fast Fourier transform on the measured interferogram to obtain the infrared absorption spectra of the samples. To improve the signal-to-noise ratio in the FTIR measurements, we transferred BP and b-AsP flakes onto a gold film to enhance their infrared reflectance, and spectra were obtained by averaging the signals of 128 scans at a spectral resolution of 2 cm$^{-1}$. In addition, the optical paths of the end-station TLS14A1 were purged by passing high-purity nitrogen gas to remove most of the water vapour and $CO_2$.

Device fabrication

The crystals of BP, b-$As_{0.25}P_{0.75}$, b-$As_{0.46}P_{0.54}$ and $MoS_2$ fakes were mechanically exfoliated onto 285-nm-thick $SiO_2$ on Si substrates by the Scotch Tape method. The thicknesses of exfoliated flakes were identified by atomic force microscopy (AFM) and the crystal orientations of exfoliated BP and b-AsP were resolved by exploiting polarization-resolved PL spectroscopy. With these identifications, we used the dry transfer technique, which allowed the precise control of the stacking sequence and relative crystallographic alignment of layered materials, to assemble the selected vdW flakes in the vertical direction to form $BP_t/MoS_2/BP_b$ or $BP/MoS_2/b$-$As_{0.46}P_{0.54}$ heterostructures. The assembled vdW heterostructures were then transferred onto a $Si/SiO_2$ substrate with two separate prepatterned Cr/Au (5 nm/30 nm) metal electrodes. During the transfer process, the bottom and top light-emitting layers were aligned to contact the two separate metal electrodes. All the above transfer and material exfoliation processes were conducted in a glove box under an inert atmosphere to minimize the oxidation of materials. The fabricated vdW heterostructure emitters were then removed from the glove box and wire bonded onto a chip carrier. After wire bonding, they were immediately loaded into the chamber and evacuated to a pressure of approximately $10^{-4}$ torr to characterize their optoelectronic properties.

Device modelling

Commercial software Lumerical DEVICE was used to simulate the energy band diagrams of $BP_t/MoS_2/BP_b$ (Fig. 2c-d) and $BP/MoS_2/b$-$As_{0.46}P_{0.54}$ (Fig. 3c-d) heterostructure emitters. These band diagrams were obtained by solving the steady state Poisson equation with the finite element method:

$$\nabla \cdot (-\varepsilon_r \varepsilon_0 \nabla V) = q(p - n + N_A - N_D)$$

where $\varepsilon_r$ is the relative permittivity; $\varepsilon_0$ is the vacuum permittivity; $n$ and $p$ are the densities of free electron and hole; and $N_A$ and $N_D$ are the acceptor and donor concentrations. The free electron and hole densities were derived by the equations:

$$n = N_C \exp(\frac{E_F - E_C}{K_B T})$$

$$p = N_V \exp(\frac{E_V - E_F}{K_B T})$$

where $N_C$ and $N_V$ are the effective density of states in the conduction and valence band; $T$ is the temperature; $E_F$ is the Fermi level; and $E_C$ and $E_V$ are the conduction and valence band energies. The related materials parameters used in the simulations were present in

Table S2 (see Supplementary Section 11), and the simulated electric field distributions of biased $BP_t/MoS_2/BP_b$ and $BP/MoS_2/b\text{-}As_{0.46}P_{0.54}$ heterostructures can be found in the Supplementary Section 11.

Frequency response characterizations

To characterize the frequency responses of emitters, we sent square-wave voltage pulses onto the heterostructure emitter. The voltage pulses were periodically varied between 0 V and 4 V (between -2 V and -10 V) to induce the modulated EL signals generated from the $BP/MoS_2$ ($b\text{-}As_{0.46}P_{0.54}/MoS_2$) unit. The modulated EL signals were then sent into an InAsSb mid-IR detector (responsivity ~4 mA/W, bandwidth ~9 MHz) and the detector signals were detected by a lock-in amplifier (Zurich Instruments, bandwidth of 50 MHz). By varying the periodicity of voltage pulses, we could resolve the change of detector signals as a function of modulation frequency as shown in Fig. 4a-b.

**Author Contributions**

**Acknowledgements:**

The authors thank Chen-Bin Huang, Kun-Ping Huang and Shang-Hua Yang for sharing electronic instruments. This work was supported by the National Tsing Hua University under grant 110QI039E1, the Ministry of Science and Technology of Taiwan under grant 107-2112-M-007-002-MY3, 109-2112-M-007-032-MY3 and 111-2124-M-007-002-MY2.

**Competing interests**

The authors declare no competing financial interests.

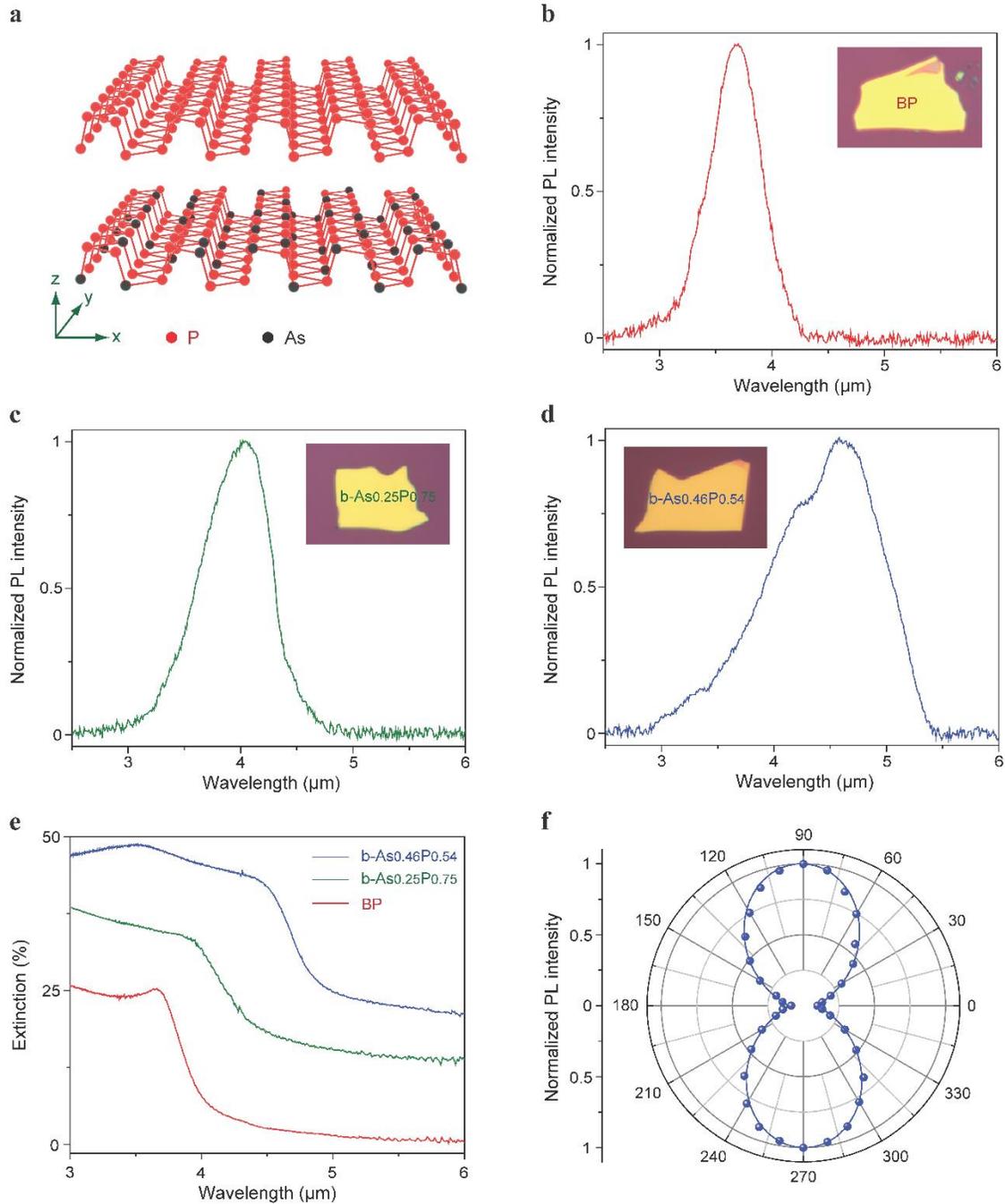

**Figure 1 Characterization of BP and b-AsP.** a. Schematic of the crystal structures of BP (top panel) and b-AsP (bottom panel). The x and y directions correspond to the armchair and zigzag directions, respectively. Red represents phosphorus atoms and black represents arsenic atoms. b-d. PL emission spectra measured from the (b) 40-nm-thick BP flake, (c) 50-nm-thick b-As$_{0.25}$P$_{0.75}$ flake and (d) 35-nm-thick b-As$_{0.46}$P$_{0.54}$ flake, respectively. All flakes are placed on SiO$_2$/Si substrates. PL spectra are slightly affected by the CO$_2$

absorption peak at $\lambda=$ ~4.3 µm. Insets: Optical images of the flakes used for PL measurements. e. Plots of extinction absorption of BP, b-$As_{0.25}P_{0.75}$ and b-$As_{0.46}P_{0.54}$. The result shows that increasing the As/P ratio causes a reduction in the band gap. The spectra are vertically shifted for clarity. We note that the absolute extinction values of the spectra cannot be compared directly, as different flakes have different thicknesses and sizes. f. Normalized integrated PL intensity of b-$As_{0.46}P_{0.54}$ as a function of the emission polarization angle, where the polarization angle of 90° (0°) corresponds to the armchair (zigzag) crystal axis. The blue line is the fitted curve to the data (blue dots) using the function $a \times \sin^2\theta + b \times \cos^2\theta$.

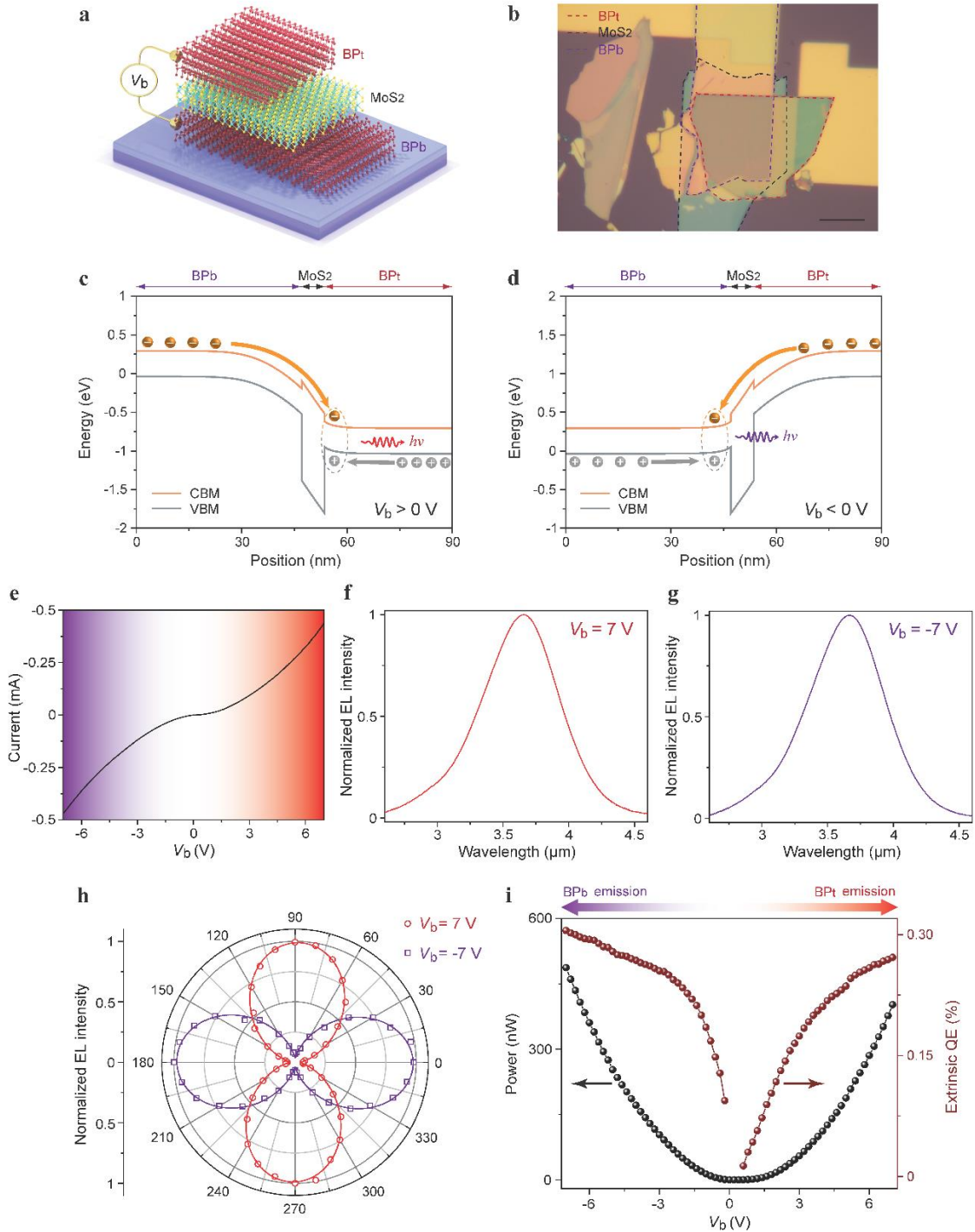

**Figure 2 Optoelectronic properties of the $BP_t/MoS_2/BP_b$ mid-IR light emitter.** a. Schematic of the device configuration, composed of vertically stacked $BP_t/MoS_2/BP_b$ heterostructures. The armchair directions of $BP_b$ and $BP_t$ are orthogonal with each other. b. Optical microscope image of the $BP_t/MoS_2/BP_b$ heterostructures. The regions of $BP_t$, $MoS_2$ and $BP_b$ are defined by red, black and purple dashed lines, respectively. Scale bar, 10 μm. c-d. Simulated energy band diagrams when applying (c) positive $V_b$, and (d) negative $V_b$

onto the $BP_t/MoS_2/BP_b$ heterostructure emitter. e. $I$-$V_b$ characteristic of the $BP_t/MoS_2/BP_b$ heterostructures. The bias is applied across two BP layers with $BP_b$ connected to the ground. f-g. Normalized EL spectra of the $BP_t/MoS_2/BP_b$ heterostructure emitter, operated at (f) $V_b$=7 V, and (g) $V_b$=-7 V, respectively. h. Normalized integrated EL intensity of the $BP_t/MoS_2/BP_b$ heterostructure emitter as a function of the emission polarization angle. The hollow red circles (purple squares) are data points, obtained by biasing the emitter at $V_b$=7 V ($V_b$=-7 V), and the solid lines are the fitted curves to the data using the function $a\times\sin^2\theta + b\times\cos^2\theta$. i. The output power (left y-axis) and extrinsic QE (right y-axis) of the $BP_t/MoS_2/BP_b$ heterostructure emitter, operated at different bias voltages.

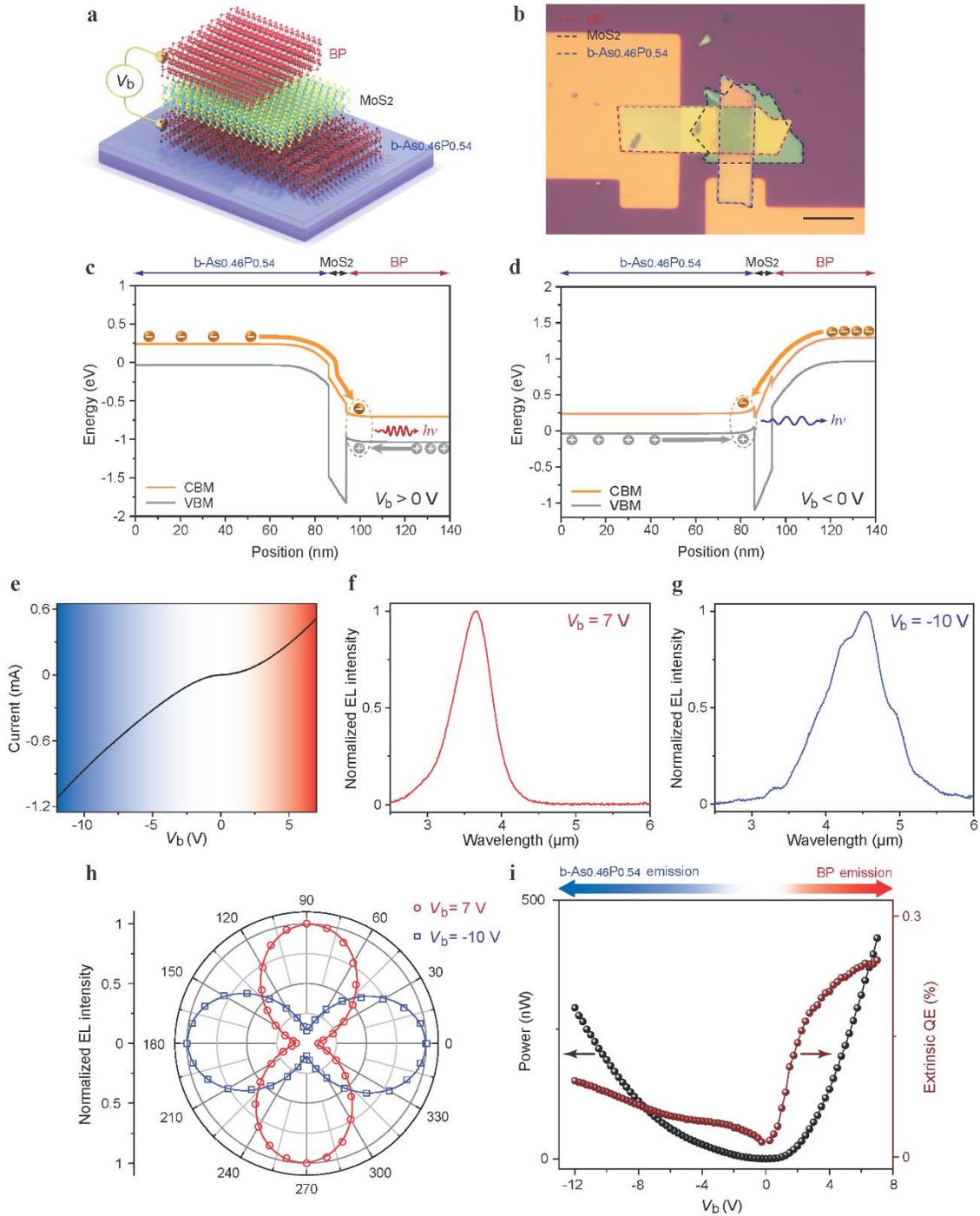

**Figure 3 Optoelectronic properties of the BP/MoS$_2$/ b-As$_{0.46}$P$_{0.54}$ mid-IR light emitter.** a. Schematic of the device configuration, composed of vertically stacked BP/MoS$_2$/b-As$_{0.46}$P$_{0.54}$ heterostructures. The armchair directions of b-As$_{0.46}$P$_{0.54}$ and BP are orthogonal with each other. b. Optical microscope image of the BP/MoS$_2$/b-As$_{0.46}$P$_{0.54}$ heterostructures. The regions of BP, MoS$_2$ and b-As$_{0.46}$P$_{0.54}$ are defined by red, black and blue dashed lines, respectively. Scale bar, 20 μm. c-d. Simulated energy band diagrams when applying (c)

positive $V_b$, and (d) negative $V_b$ onto the BP/MoS$_2$/b-As$_{0.46}$P$_{0.54}$ heterostructure emitter. e. $I-V$ characteristic of the BP/MoS$_2$/b-As$_{0.46}$P$_{0.54}$ heterostructures. The bias is applied across b-As$_{0.46}$P$_{0.54}$ and BP layers with b-As$_{0.46}$P$_{0.54}$ connected to the ground. f-g. Normalized EL spectra of the BP/MoS$_2$/b-As$_{0.46}$P$_{0.54}$ heterostructure emitter, operated at (f) $V_b$=7 V, and (g) $V_b$=-10 V, respectively. h. Normalized integrated EL intensity of the BP/MoS$_2$/b-As$_{0.46}$P$_{0.54}$ heterostructure emitter as a function of the emission polarization angle. The hollow red circles (blue squares) are data points, obtained by biasing the emitter at $V_b$=7 V ($V_b$=-10 V), and the solid lines are the fitted curves to the data using the function $a \times \sin^2\theta + b \times \cos^2\theta$. i. The output power (left y-axis) and extrinsic QE (right y-axis) of the BP/MoS$_2$/b-As$_{0.46}$P$_{0.54}$ heterostructure emitter, operated at different bias voltages.

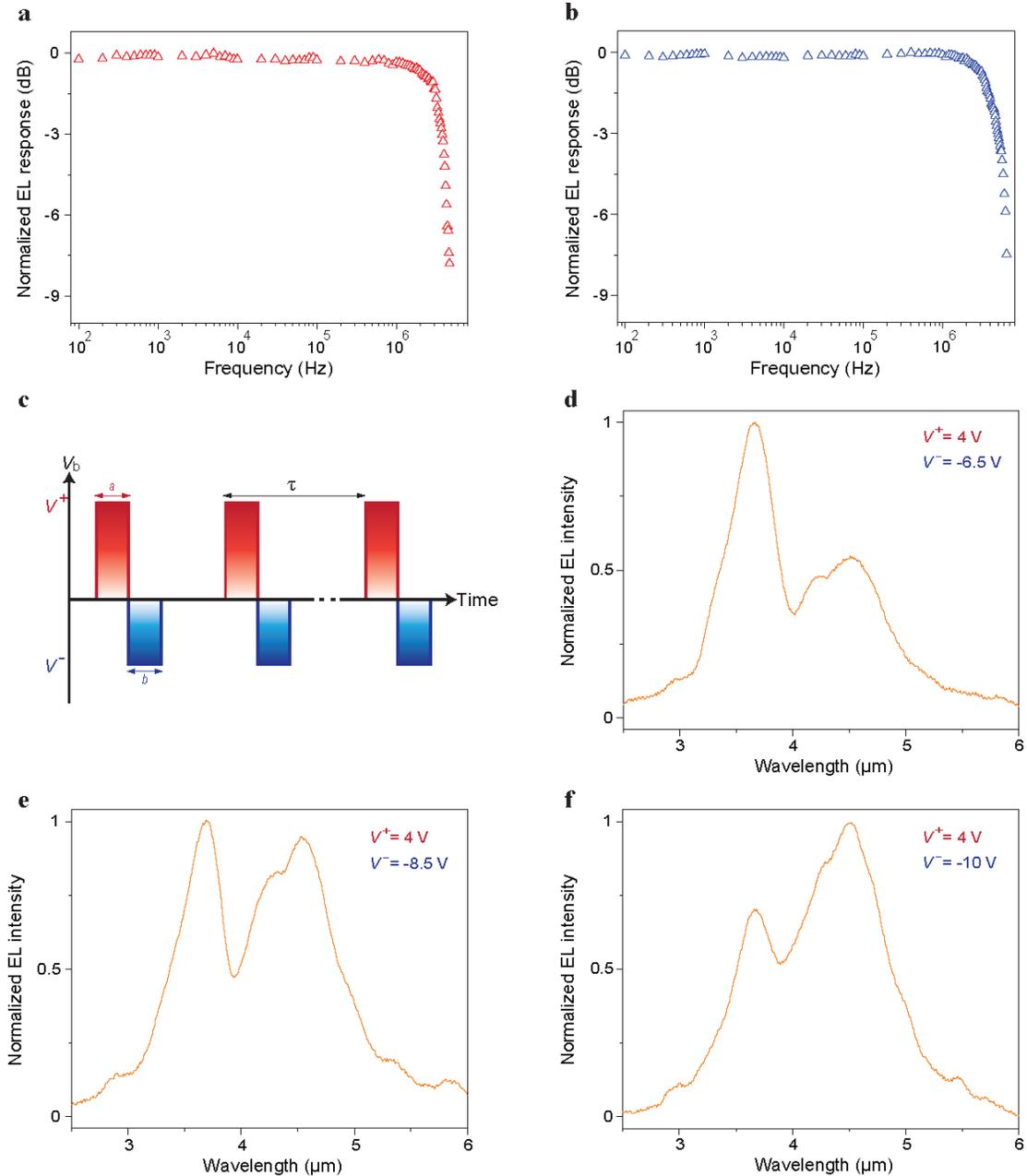

**Figure 4 Pulsed operation of the BP/MoS$_2$/b-As$_{0.46}$P$_{0.54}$ mid-IR light emitter.** a-b. Frequency responses of the BP/MoS$_2$/b-As$_{0.46}$P$_{0.54}$ heterostructure emitter. The applied bias voltage periodically varies between (a) 0 V and 4 V, and (b) -2 V and -10 V at different frequencies. c. Schematic illustration of the applied voltage pulse sequence. $V^+$: Peak voltage of positive pulses. $V^-$: Peak voltage of negative pulses. $\tau$: Pulse period. $a$: Time interval of the positive pulse. $b$: Time interval of the negative pulse. d-f. Normalized EL spectrum when applying voltage pulse trains onto the BP/MoS$_2$/b-As$_{0.46}$P$_{0.54}$ heterostructure emitter with (b) $V^+$ set to 4 V and $V^-$ set to -6.5 V, (c) $V^+$ set to 4 V and $V^-$

set to -8.5 V, and (d) $V^+$ set to 4 V and $V^-$ set to -10 V. For all experiments, $\tau$ is set to 1 ms, and the duty cycles of the positive and negative pulses are both set to 15%.